\documentclass[floats,preprint,prl,aps]{revtex4}
\usepackage{latexsym}
\usepackage{graphicx}
\newcommand\beq{\begin{equation}}
\newcommand\eeq{\end{equation}}
\newcommand\bea{\begin{eqnarray}}
\newcommand\eea{\end{eqnarray}}
\newcommand\non{\nonumber}
\newcommand\bib{\bibitem}
\begin{document}
%\textheight=24cm
%\twocolumn[\hsize\textwidth\columnwidth\hsize\csname@twocolumnfalse\endcsname
\title{\bf A Renormalization Group Study of Interacting Helical Liquid  
: Physics of Majorana Fermion}

\author{\bf Sujit Sarkar}
\address{\it Poornaprajna Institute of Scientific Research,
4 Sadashivanagar, Bangalore 5600 80, India.\\
}
\date{\today}

\begin{abstract}
The physics of Majorana fermion occurs at the edge and interface of many 
one-dimensional quantum many body system with gapped excitation spectrum.
We present the  
renormalization group
equations for strongly interacting helical liquid. 
We present the results of both Majorana-Ising topological
quantum phase transition and also Berezinskii-Kosterlitz-Thouless
topological phase transition. 
We show that the
Majorana-Ising topological quantum phase transition for
non-interacting case is an exactly solvable line but 
in presence of interaction the system has no exactly
solvable line. We study the effect of umklapp scattering on the
renormalization group flow diagrams and also the condition for
the appearence of Majorana fermion modes.
\\

% Pacs: 42.50.Dv, 42.50.Pq, 03.67.Bg, 75.10.Jm \\  
\\
\end{abstract}
%\vskip .5 true cm
\maketitle

\section{I. Introduction}
%\vskip -1.5 true cm
%{\bf Introduction:} 
The physics of Majorana fermion is the subject of intense research interest
in quantum many body system near a deca [1-31].
Majorana fermion appeares in the topologically ordered
state which is one of the important focous of research in condensed matter
physics. These states are chacterized by the topological invariant quantities
, Chern number and Winding number. 
Kitaev \cite{kitaev} and the other research group have proposed the physical realization
of Majorana fermion as a edge state of the of one dimensional system
that include one-dimensional superconductors, semiconductor quantum wire
, proximity coupled to superconductor, the cold atom trapped in
one-dimension. The physics and the control technology of Majorana fermion are
the most important properties of non-abelian quantum computation,
the quantum Hall states are the example of topological states \cite{nayak}. 
Therefore the physics of Majorana fermion is almost ubiquitous
in different quantum many body system [1-31].\\
In this research paper, 
we study for the existence of Majorana fermion mode in interacting helical liquid  
by using the renormalization group theory method. \\
Before we proceed further we would like
to state the basic properties of Helical liquid very briefly [9-12,16-31].
The physics of helical liquid is interesting for the 
following properties. 
It is generally originated from the quantum spin Hall effect in a system
with or with out Landau levels. In the quantum spin Hall effect the left
movers in the edge are connected with the down spin and the right movers
with the up spin and the transport process is quantized. This physics is
generally termed as a "Helical liquid" which describe the connection between
the spin and momentum. It does not break time reversal invariance which
occurs in chiral Luttinger liquid. 
The
helical liquid has even number of time reversal pair whereas the spinful
Luttinger liquid has odd number of pairs. The physics of helical liquid
occurs in the surface state of a two dimensional insulator a proximity
coupled semiconductor quantum wire subject to the spin orbit interaction. \\
Here we describe the basic mathematical analysis of Helical liquid. 
It is well known to us that the low energy excitation in the one dimensional
quantum many body system occurs near to the adjacent region of Fermi points.
Therefore one can write the fermionic field operator as \cite{gia}
\beq
{\psi}_{\sigma} (x) = \frac{1}{\sqrt{L}} [ \sum_{-\Lambda < k - k_F < \Lambda}
e^{i k.x} {\psi}_{\sigma} + \sum_{-\Lambda < k + k_F < \Lambda}
e^{i k.x} {\psi}_{\sigma} ] . 
\eeq 
Where $\Lambda $ is the cut-off around the Fermi momentum ($k_F $).
We may consider the first term as a right mover ($ k>0 $) and the second term
as a left mover ($ k < 0 $). One can write the fermionic field with spin 
$\sigma $ as 
$ {\psi}_{\sigma} (x) = {\psi}_{R \sigma} (x) + {\psi}_{L \sigma} (x) $.
For the low energy elementary excitations one can write the Hamiltonian
as \\
\bea
H_0 &=& \int \frac{dk}{2 \pi} {v_F} [ {{\psi}_{R \uparrow}}^{\dagger} (i \partial_x)
{\psi_{R \uparrow} } - i {{\psi}_{L \downarrow}}^{\dagger} (i \partial_x)
{\psi_{L \downarrow} } \non \\
& & +  {{\psi}_{R \downarrow}}^{\dagger} (i \partial_x)
{\psi_{R \downarrow} } - i {{\psi}_{L \uparrow}}^{\dagger} (i \partial_x)
{\psi_{L \uparrow} } ] .
\eea
The first term within the first bracket is one of the Kramer's pair and
the second term within the first bracket is the another Kramer's pair.
Therefore one of the Kramer's pair in the upper edge and the other one
at the lower edge of the system. The total fermionic field,  
${\psi} (x) = {\psi}_{R \uparrow} + {\psi}_{L \downarrow} $, 
which can be expressed as a spinor 
$ \psi = {( {\psi}_{L \downarrow} ~~ {\psi}_{R \uparrow} )}^{T} $. 
This is the simple mathematical picture of a helical liquid,
where the spin is determined by the direction of the particle. 
The non-interacting part of the helical liquid for single edge
in terms of spinor field is
\beq
H_0 = {\psi}^{\dagger}  (i v_F \partial_x  {\sigma}^z  - \mu) {\psi} .
\eeq 
The authors of Ref.\cite{sela} have introduced a model Hamiltonian to study
the Majorana fermion in strongly interacting helical liquids. 
Here we describe the model very briefly of Ref. \cite{sela}.
\\
The authors have described a low-dimensional quantum many body
system of topological insulator in the proximity of s-wave superconductor
and an external magnetic field along the edge of this system. 
The additional terms in 
the Hamiltonian is 
\beq
\delta H =  \Delta {\psi_{L \downarrow}} {\psi_{R \uparrow}} + 
B {{\psi}_{L \downarrow}}^{\dagger}{\psi_{L \uparrow}} + h.c . 
\eeq
Where $\Delta $ is the proximity induce superconducting gap and $B$ is
the applied magnetic field along the edge of the sample. The Hamiltonian
$H_0 $ is the time reversal invariant but the Hamiltonian $\delta H$ is
not time reversal invariant.\\
Now we consider the generic interaction which are time reversal, the authors
of Ref. \cite{sela}, have considered the two particle forward and umklapp scattering
as 
\beq
H_{fw} = g_2  {\psi_{L \downarrow}}^{\dagger} {\psi_{L \downarrow}}
 {\psi_{R \uparrow}}^{\dagger} {\psi_{R \uparrow}} . 
\eeq
In the umklapp scattering term, we write the umklapp term for
half filling following the Wu, Bernevig and Zhang \cite{wu}, the umklapp
term at the half filling in a point splitted form. 
The point splited version can be described as a regularization 
of the theory.
Therefore the
umklapp term become 
\beq
H_{um} = - g_u \int dx 
{{\psi}_{R \uparrow}}^{\dagger} (x) 
{{\psi}_{R \uparrow}}^{\dagger} (x + a) 
{{\psi}_{L \downarrow}}^{\dagger} (x) 
{{\psi}_{L \downarrow}}^{\dagger} (x + a) .  
\eeq 
Where $a$ is the lattice constant. This analytical 
expression gives a regularized theory using the lattice
constant $a$ as a ultraviolet cut-off.\\
If we use the first order Taylor series expansion of the fermionic
field 
\beq 
{{\psi}_{R \uparrow}}^{\dagger} (x + a) 
\sim {{\psi}_{R \uparrow}}^{\dagger} (x)  + a \partial_x 
{{\psi}_{R \uparrow}}^{\dagger} (x) .
\eeq 
If we use this expansion
for the umklapp scattering term then we produce the analytical
expression for umklapp in a conventional 
form of the authors of Ref.\cite{wu}.
\beq
H_{um} = g_u {{\psi}_{L \downarrow}}^{\dagger} \partial_x {{\psi}_{L \downarrow}}^{\dagger}
{{\psi}_{R \uparrow}} \partial_x {{\psi}_{R \uparrow}} + h.c
\eeq
Therefore the total Hamiltonian of the system is
\beq
H = H_{0} + H_{fw} + H_{um} + \delta H .
\eeq 
Now we can write the above Hamiltonian as,
$H_{XYZ} = \sum_i H_i $ (up to a constant) \cite{sela}. \\
\beq
H_i = \sum_{\alpha} J_{\alpha} {S_i}^{\alpha} {S_{i+1}}^{\alpha}
- [ \mu + B (-1)^i ] {S_i}^z  .
\eeq
Where $J_{x,y} = J \pm \Delta >0 $ and $J =v_F $ and $J_z  >0 $. \\
Here we write down the bosonized form of this model Hamiltonian, the
detail derivation is relegated to appendix.\\
\bea
H & = & \frac{v}{2} ( \frac{1}{K} ( {({\partial_x \phi })}^2 + 
K {({\partial_x  \theta })}^2  ) - (\frac{\mu}{\pi}) \partial_x \phi \non\\
& & + \frac{B}{\pi} cos(\sqrt{4 \pi} \phi) - \frac{\Delta}{\pi} cos(\sqrt{4 \pi \theta})
+ \frac{g_u}{2 {\pi}^2 } cos(4 \sqrt{\pi} \phi) 
\eea
There is a sign mismatch between the present study and sign of Ref. \cite{sela}, please
see the appendix for detail explanation.\\
The author of Ref. \cite{sela} have studied the quantum phase diagram
of this model Hamiltonian based on the Abelian bosonization method.
But here we study the same Hamiltonian by using the RG method for the
following reasons. 
It is very clear from the continuum field theoretical study 
that our model Hamiltonian contains two 
strongly relevant and mutually nonlocal perturbation over
the Gaussian (critical) theory.
In such a situation the strong coupling fixed point is usually
determined by the most relevant perturbation whose amplitude
grows up according to its Gaussian scaling dimensions and
it is not much affected by the less relevant coupling terms.
However, this is not the general rule if the two operators
exclude each other, i.e., if the field configurations which
minimize one perturbation term do not minimize the other.
In this case interplay between the two competing relevant
operators can produce a novel quantum phase transition through
a critical point or a critical line. Therefore, we would like to
study the RG equation to interpret the quantum phases of the system.\\
%%%%%%%%%%%%%%%%%%%%%%%%%%%%%%%%%%%%%%%%%%%%%%%%%%%%%%%%%%%%%%%%%%%%%%%%%%%%%%%%%% 
\section{ Analysis of Renormalization Group Equations} 
We now study how the parameters $B$, $\Delta$ and $K$ flow under RG.
through the
operator product expansion. So the RG equations for their coefficients
therefore are coupled to each other.
We use operator product expansion to derive
these RG equations which is independent of boundary condition
\cite{suj}. 
In our derivation, we 
consider two operators,
$ X_1 = e^{(i a_1 \phi + i b_1 \theta)}$ and $ X_2 = e^{(i a_2 \phi + i b_2 \theta)}$.
In the RG procedure, one can write these two field operators as a sum of
fast and slow mode fields. In the fast field, the momentum range is 
$ \Lambda e^{-dl} < K < \Lambda $ and for the slow field 
$ K < \Lambda e^{-dl} $,
where $\Lambda $ is the momentum cut-off, $dl $ is the change in the logarithmic
scale. The next step is the integration of the fast field for the operators $X_1 $
and $X_2 $, it yields a third operator at the same space time point,
$ X_3 = e^{i (a_1 + a_2) \phi + i (b_1 + b_2) \theta)}$. The prefactor of $X_3 $
can be found by the relation, 
$ {X_1 }{X_2} \sim e^{-(a_1  a_2 + b_1 b_2 )} \frac{dl}{2 \pi} X_3 $.
Our Hamiltonian consists of two operators, 
if we consider $l_1 $ and $l_2 $ as the coefficient of
the operators $X_1 $ and $X_2 $ respectively. Then the RG expressions for $\frac{d X_3}{dl}$
contains the term $ (a_1 a_2 + b_1 b_2 ) \frac{l_1 l_2}{2 \pi} $. This is the procedure
to derive these RG equations.\\
In the RG process, one can write 
RG equations themselves are established
in a perturbative expansion in coupling constant ($g(l)$). They
cease to be valid beyond a certain length scale, where
$g(l) \sim 1$ \cite{gia}.
\bea
\frac{dB}{dl} ~&=&~ (2 - K ) B + 4 K g_u B ,\non \\
\frac{d \Delta}{dl} ~&=&~ (2 -\frac{1}{K}) a ,\non \\
\frac{d{g_u}}{dl} ~&=&~ (2 - 4K) g_u + 2 K B^2  \non \\
\frac{dK}{dl} ~&=&~ \frac{a^2}{4} ~-~ K^2 {\Delta}^2 ~.
\label{rg2}
\eea
Before we proceed further to study the detail analysis of these
four RG equations. Here at first we consider the absence of
umpklapp scattering, therefore the four RG equations reduced to
three RG equations. 
The RG equations for the coefficients of Hamiltonian $H_{XYZ}$ are
\bea
\frac{d \Delta}{dl} ~&=&~ (2 -\frac{1}{K}) \Delta ,\non \\
\frac{d{B}}{dl} ~&=&~ (2 - K) {B}  \non \\
\frac{dK}{dl} ~&=&~ \frac{\Delta^2}{4} ~-~ K^2 {B}^2 ~.
\label{rg2}
\eea
Here our main intention is to study these three equations explicitly
to explain the different quantum phases of the system and also the
Majorana-Ising quantum phase transition.
After few steps of calculation, one can also write the above three RG equations 
as  
\bea
\frac{d \Delta}{dl} ~&=&~ (2 -\frac{1}{K}) \Delta ,\non \\
\frac{d{B}}{dl} ~&=&~ (2 - K) {B}  \non \\
\frac{dlnK}{dl} ~&=&~ {K}^{-1} {\Delta^2} ~-~ K {B}^2 ~.
\label{rg2}
\eea
where $ l= ln[\frac{\Lambda}{\Lambda_0}] $ is the flow parameter, 
$ {\Lambda_0 }$ is the initial value of the parameter.
It is very clear from the above RG equations which reflect the duality
in our helical liquid model system. The duality is the following.
$ \phi \leftrightarrow \theta$, $ K \leftrightarrow {K}^{-1} $ and
$ \Delta \leftrightarrow B $. This duality in RG equations is actually 
from the duality, i.e., if we interchange the $B$ and $\Delta$ and
simultaneously the field $\phi$ and $\theta$ then the Hamitonian
remains invariant of the bosonized Hamiltonian.\\  
These RG equations have trivial (${\Delta^*}= 0 = {{B}^*}$)
fixed points for any arbitrary $K$. 
Apart from that these RG equations have also two non-trivial
fixed lines, $\Delta = B $ and ${{\Delta} } =- B$
for $ K =1 $.\\
It is very clear from the above RG equations, in the vicinity of $K =1$,
one can find the non-trivial critical points, where both of the coupling 
term are become the relevant coupling, i.e.,the coupling constant flowings
off to the strong coupling limit, where each of them acquires a mass term.\\
The coupling strengths $\Delta $ and $B$ are related to the dual fields,
$\theta$ and $\phi$ that drives the system to the different ground state,
one is in the proximity induced superconducting gap and the other is in
the applied magnetic field induce ferromagnetic phase \cite{sela,wu}. 
The interplay between them
give rise to the second order quantum phase transition at the intermediate couplings.
We have already discussed that the $K=1$, the self dual point and this
point is also the only exactly solvable point.\\ 
In our RG calculation, we also predict the Majorana-Ising transition
. Before we proceed further
we would like to explain the basic aspect of Majorana-Ising quantum
phase transition.\\
\beq
\delta H = B {{\psi}_{L \downarrow}}^{\dagger} {\psi}_{R \uparrow}  + 
\Delta {\psi}_{L \downarrow} {\psi}_{R,\uparrow} + h.c .
\eeq
We recast the fermionic field interms of the Majorana fields,
$ {\psi}_{L \downarrow} (x) = \frac{1}{2} (i \chi_1 (x) + \chi_2 (x) ) $. 
$ {\psi}_{R \uparrow} (x) = \frac{1}{2} ( {\tilde{ \chi}}_1 (x) 
+ i {\tilde{ \chi}}_2 (x) ) $.
The total Hamiltonian, $ H = H_{0} + \delta H $, become
\beq
H = \sum_{i=1,2} ( i {\chi}_i \frac{v_F}{2} \partial_x {\chi}_i -
i \tilde{{\chi}_i } \frac{v_F}{2} \partial_x \tilde{{\chi}_i} 
+ i m_i {{\chi}_i} \tilde{{\chi_i}}). 
\eeq  
Where $m_{1,2} = \Delta \mp B $ (here $\Delta > 0$ ).
At $\Delta = B $, one of the two Majorana fermion modes
become gapless which is the signature of bulk Majorana-Ising 
quantum phase transition. We will explore this transition explicitly
during the renormalization group study.\\
Now we study the scaling theory for this problem.
It is well known to all of us that the critical theory is invariant under
the rescaling, the singular part of the free energy density satisfies the
following scaling relations.\\
\beq
 f_s [\Delta, B] = e^{-2l} f_s [e^{(2 - 1/K)l} \Delta , e^{(2 -K )l
} B] .
\eeq 
The scale $ l$ can be fixed from the following analytical relation,
$ e^{(2 -1/K ){l}^{*} } \Delta =1 $. After a few substituation, we
arrive at the follwing relation.
\beq
f_s [\Delta, B] = {\Delta}^{2/(2 - 1/K)} f_s [1 , {\Delta}^{-(2-K)/(2 -1/K)} B] .
\eeq
The phase transition of the system occurs when the coupling strength
satisfy the following relation:\\
\beq
{\Delta}^{-(2-K)/(2 -1/K)} B \sim 1
\eeq
The phase boundary between this two quantum derive by using the above relation.
When $K=1$, (non-interacting case), the above phase boundary 
relation is $\Delta =B$, which we have already discuss in the previous section,
when we introduce the basic physics of Majorana-Ising topological quantum phase
transition.\\
Now we discuss, the effect of repulsive ($ K < 1$) and attractive ($ K > 1$) region
of this phase when the umklapp interaction is present. These results are depicted
in the fig.1. It observes from our study that the phase bounday
exists for the non-interating and also for the interacting one. The phase bounday
is a exactly solvable line for the non-interacting case but for the interacting
case
it is not exactly solvable line.\\ 
One can understand this shift of this phase boundary from 
the following mathematical analysis.
Generally one can consider the interaction 
between the Majorana fermions as
$ H_{int} \sim {\chi}_1 \bar{\chi_1} {\chi}_2 \bar{\chi_2 } $. 
In the mean field level, one can do the following approximation,
$ {\chi}_2 \bar{\chi_2} \rightarrow < \chi_2 \bar{\chi_2} > $. Thus it is clear from
the mean-field analysis that one can absorab the effect of interaction as a redefine
mass in the system (Eq.16), which is shifting the phase boundary.\\
Here we discuss very briefly based on this toy model about the nature of
quantum phases during the topological quantum phase transition.\\
In our toy model, we consider the simple case, i.e., the umklapp
term is absent and at the point $ \Delta = J$. The complete Hamiltonian
reduce to transverse Ising model.\\
\beq
H_i = 2 \Delta {s_i}^{x} {s_{i+1}}^{x} - B {s_i}^{z} 
\eeq   
when $ B < \Delta $, the discrete Ising symmetry is spontaneously
broken which yields a doubly degenerate ordered phase which is
proximity effect induce superconducting gap state which form
the Majorana fermion mode excitations at the edge of the system.
For $ B > \Delta $, the magnetic field induce ferromagnetic 
state along the direction of magnetic field. \\
\begin{figure}
\includegraphics[scale=0.7,angle=0]{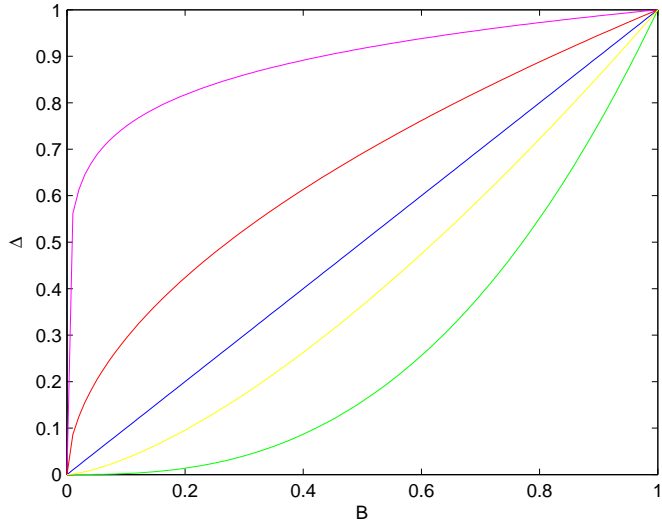}
\caption{ (Colour online.) This figure presents the phase boundary between the
two different phases $\Delta$ and $B$. 
The color blue,magenta,red,green and yellow for $ K = 1, 0.55,0.75, 1.5, 1.2 $  
respectively. The blue line is the exactly solvable line. }
\label{Fig. 1 }
\end{figure}
\begin{figure}
\includegraphics[scale=0.7,angle=0]{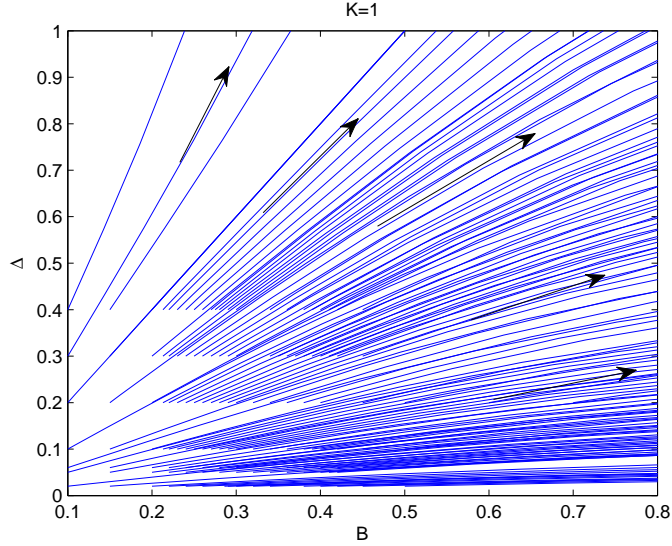}
\caption{ (Colour online.) This figure shows the RG flow of $\Delta $ with B
for non-interacting helical liquid, i.e, $K = 1$ inabsence
of umklapp scattering. 
}
\label{Fig. 2 }
\end{figure}
\begin{figure}
\includegraphics[scale=0.7,angle=0]{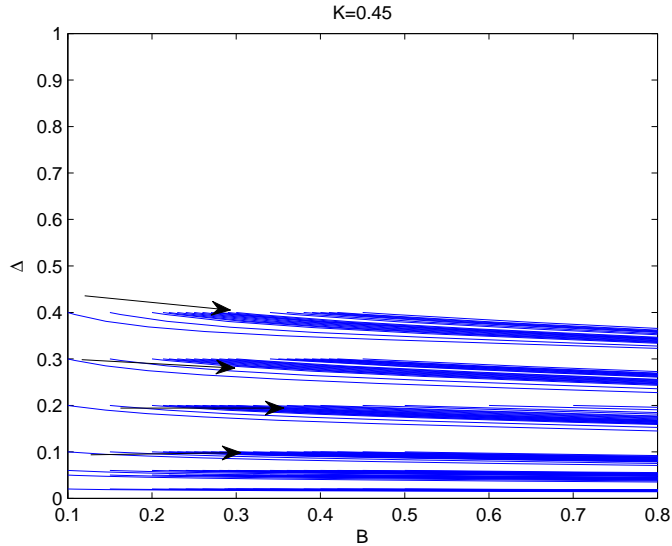}
\caption{(Colour online.) This figure shows the RG flow lines of $\Delta $
with $ B$ for interacting helical liquid in absence of
umklapp scattering. Here $ K =0.45 $ and $ B= 0.2$. 
}
\label{Fig. 3 }
\end{figure}
\begin{figure}
\includegraphics[scale=0.7,angle=0]{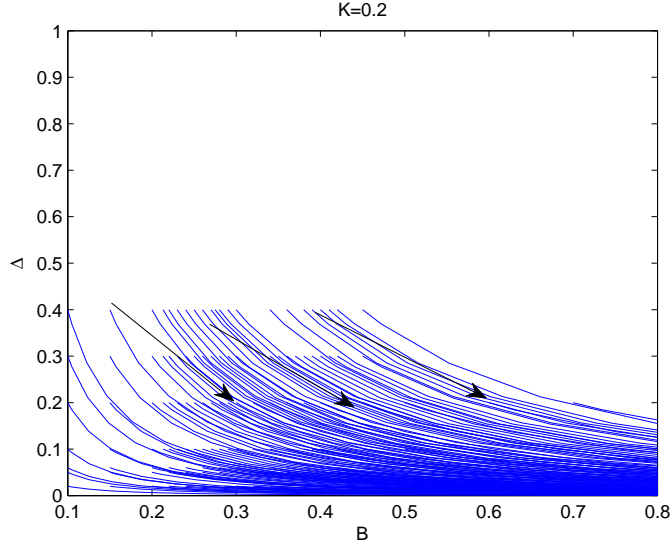}
\caption{(Color online.) This figure shows the RG flow lines of
$\Delta $ with $B$ for strongly repulsive 
regime where $K =0.2$ in absence of umklapp scattering.  
}
\label{Fig. 4 }
\end{figure}
\begin{figure}
\includegraphics[scale=0.7,angle=0]{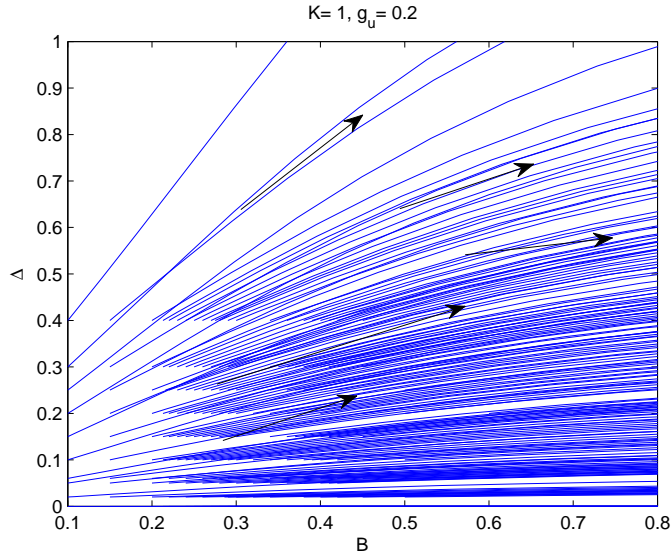}
\caption{(Colour online.) This figure shows the RG flow lines of $\Delta$
with $B$ for $ K=1 $ in presence of umklapp scattering, $g_u =0.2 $.
Here $B =0.2$.   
}
\label{Fig. 5 }
\end{figure}
\begin{figure}
\includegraphics[scale=0.7,angle=0]{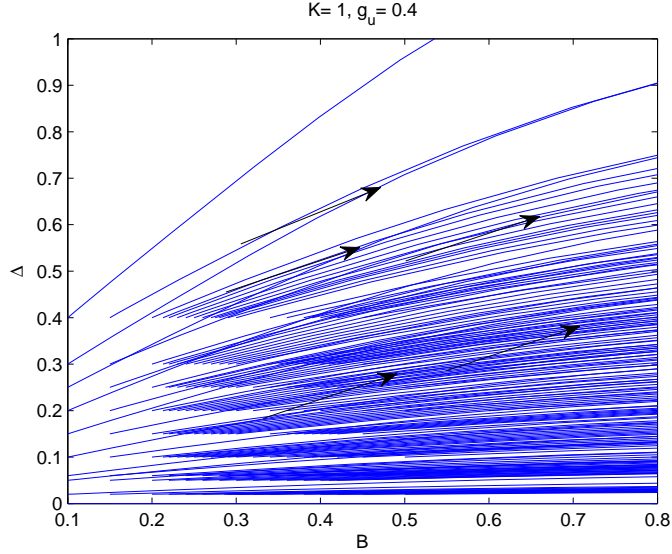}
\caption{(Colour online.) This figure shows the RG flow lines of $\Delta$ 
with $B$ for $ K= 1$ in presence of umklapp scattering, $g_u =0.4 $.
}
\label{Fig. 6 }
\end{figure}
\begin{figure}
\includegraphics[scale=0.7,angle=0]{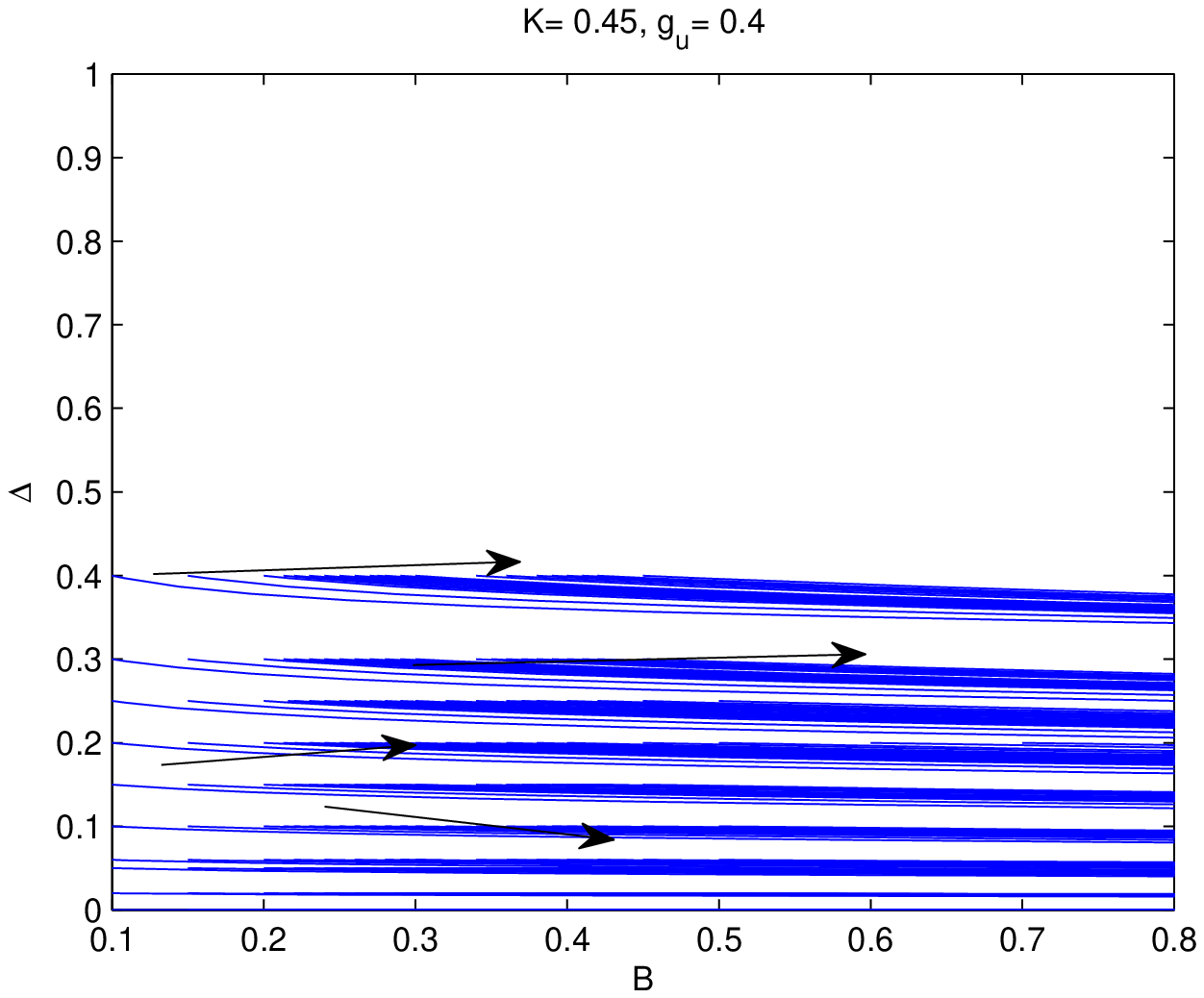}
\caption{(Colour online.) This figure shows the RG flow lines of $\Delta$ 
with $B$ for $K =0.45 $ and $g_u =0.4 $. 
}
\label{Fig. 7 }
\end{figure}
%%%%%%%%%%%%%%%%%%%%%%%%%%%%%%%%%%%%%%%%%%%%%%%%%%%%%%%%%%%%%%%%%%%%%%%%%%%%%%%%%%%%%%
\section{ Results and Discussions Based on RG Equations Study }
In fig.2, we present the RG flow diagram for $\Delta$ with $B$ for $K=1$. 
It reveals from our study that
both the couplings ( $\Delta$ and $B$) are flowing off to the strong coupling
phase. Here both the coupling terms are relevant, i.e., both of the
coupling terms flowing off to the strong coupling phase.\\
In fig.3, we present the RG flow diagram for $\Delta $ with $B$ for $K=0.45$.
For this value of $K$, system is in the repulsive regime. According to the
analysis of Abelian bosonization, the coupling term $\Delta $ is irrelevant 
but the RG flow diagram of our study shows that the $\Delta $ is non-zero
but the magnitude has reduced. This is the advantage of the study of RG
flow diagram over the Abelian bosonization study. 
It is also clear from this study that for this value of $K$, the magnitude of
$\Delta$ is almost the same it's initial value, i.e., the values of $\Delta$s
are not changing during the RG flow.
The proximity induce
superconducting gap is non-zero, i.e, the Majorana fermion mode exists 
at the edge of topological insulator of the system.\\
In fig.4, we present the RG flow diagram $\Delta$ with $B$ for $ K=0.2 $, i.e, 
the system is in the strongly repulsive regime compare to the $K =0.45 $. It reveals
from our study that the coupling term, $\Delta$, flowing off to zero, i.e.,
for this case there is no existence of Majorana fermion mode in the system. The
magnetic field induce ferromagnetic phase dicted by the direction of magnetic field
is the only phase exist in the system.\\
Therefore, it is clear from the study of fig.2, fig.3 and fig.4 that
the Majorana fermion mode only disappear in presence of very strong repulsion
, otherwise, Majorana fermion mode is robust for weak and intermediate values
of strong repulsion.\\  
In fig.5 and fig.6, we present the results for $g_u =0.2 $ and $g_u =0.4 $ respectively
for $K= 1$. It is very clear that umklapp scattering unable to destroy the superconducting
phase,i.e., the existence of Majorana phase is always in the system. The qualitative behavious
is the same for the both values of umklapp scattering.\\
In fig.7, we present the result of the variation of $\Delta$ with $B$ for $K=0.45$ and
$ g_u =0.4 $. It is clear to us that the behaviour of the system for repulsive regime is
the same in presence of umklapp scattering also. The magnitude of the proximity induce
superconductivity gap deceeasecs but it is nonzero, therefore the presence of Majorana
fermion mode is still exists.\\     
In the present situation, our system is 
topological insulator in the vicinity of an s-wave superconductor and an
external magnetic field. Therefore the appearence of the topological phase, i.e.,
the appearence of Majorana fermion mode at the edge of the topological insulator
depends on the proximity induce gap in the system. Here we study, the
length scale dependent of the system and show how it appears. The survival condition
for the Majorana fermion mode at the edge of the system is the following. 
If the superconducting region
has a length $L $, the Majorana mode survive if the condition, $L \Delta >>1 $,
otherwise there is no Majorana fermion mode, i.e., the topological state of the
system is absent \cite{suhas}. \\
Therefore, it is clear from the study fig.5, fig.6 and fig.7 that the presence
of umklapp scattering does not affect the appearence of Majorana fermion mode.\\
In fig.8, we present the results of the study of $\Delta$, $B$ and $K$ with the
length scale ($ l $). Here we consider three values of $\Delta= 0.1, 0.2, 0.4$
as the three initial values of $\Delta$ but the initial value of $B= 0.2$ and
$ K= 1$. It is clear from this length scale dependent study that the condition
for the topological state of matter,i.e., the existence of Majorana fermion
mode at the edge of the system.\\ 
In fig.9, we present the results of the study of 
$\Delta$, $B$ and $K$ with the length scale under RG process. Here
we consider $K= 0.45$, i.e, the repulsive regime of the system. It is clear
from our study that $B$ is increasing very rapidtely with length scale. 
But for the higher values of $\Delta$s the length scale dependent
is almost constant or decreases very slowly. 
It is clear from this figure 
that for the smaller values of $\Delta$s 
decreasing very rapidly with length scale
and the condition for Majorana fermion mode existence  
is not satisfied for the large values of $\Delta$, the condition for 
Majorana fermion mode 
satisfies \cite{suhas}, 
and the system is in the topological state. \\ 
In. fig. 10, we present the variation for three different values of $\Delta$s
with length under the RG process. The other parameters of the system are,
$ K=0.2$ and $B=0.2$. For this parameter space, the system is in the strong 
repulsive phase and the proximity induce superconducting gap goes to zero value
very rapidtely. In this
parameter space there is no evidence of Majorana fermion mode in the system.\\ 
 
\begin{figure}
\includegraphics[scale=0.8,angle=0]{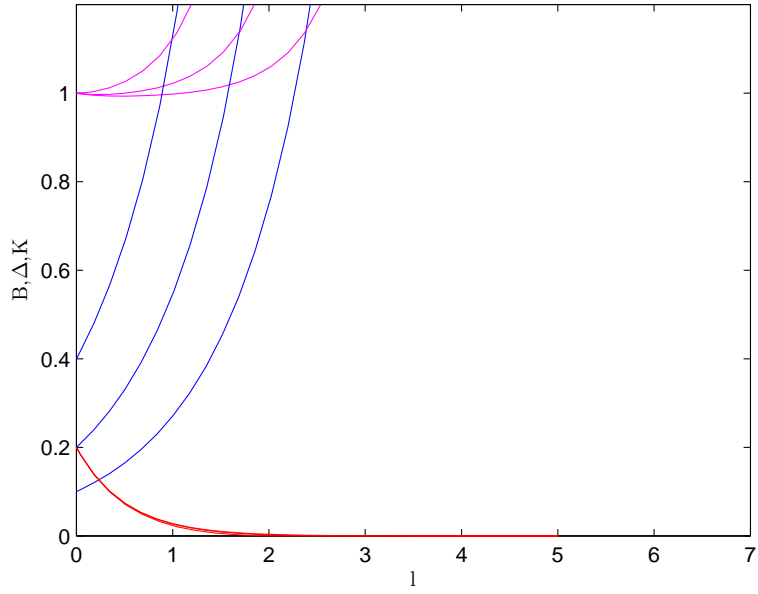}
\caption{ 
(Colour online.) This figure shows the  
variation of $\Delta, B$ and $K$ with length scale.
The curves for blue, pink and red lines are 
respectively for the $\Delta$, $K$ and $B$.  
$ \Delta = 0.1, 0.2, 0.4;  B=0.2, K=1 $,
}
\label{Fig. 8 }
\end{figure}
\begin{figure}
\includegraphics[scale=0.8,angle=0]{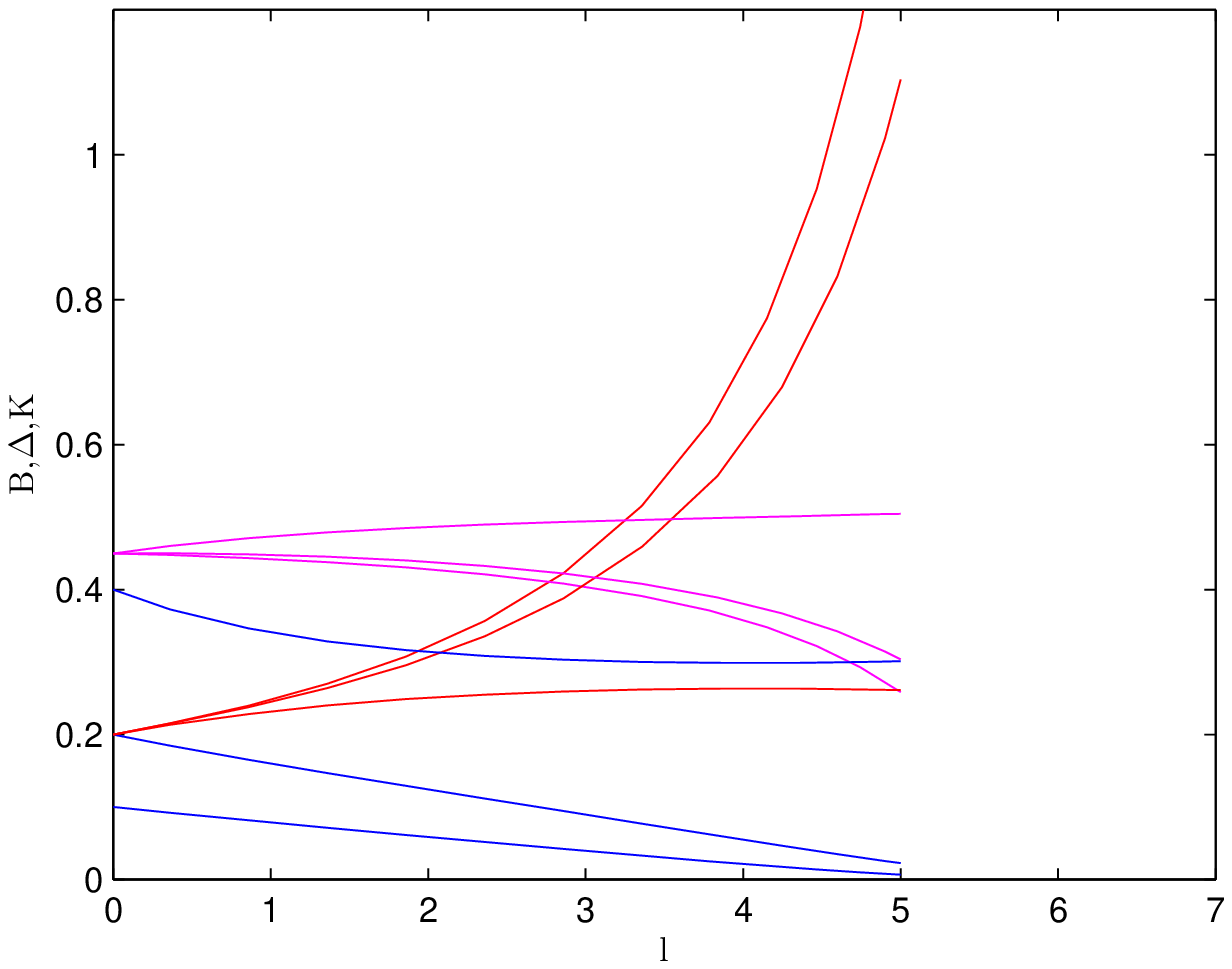}
\caption{  
(Colour online.) This figure shows the  
variation of $\Delta, B$ and $K$ with length scale.
The curves for blue, pink and red lines are 
respectively for the $\Delta$, $K$ and $B$.  
$ \Delta = 0.1, 0.2, 0.4; B=0.2, K= 0.45 $,
}
\label{Fig. 9 }
\end{figure}
\begin{figure}
\includegraphics[scale=0.8,angle=0]{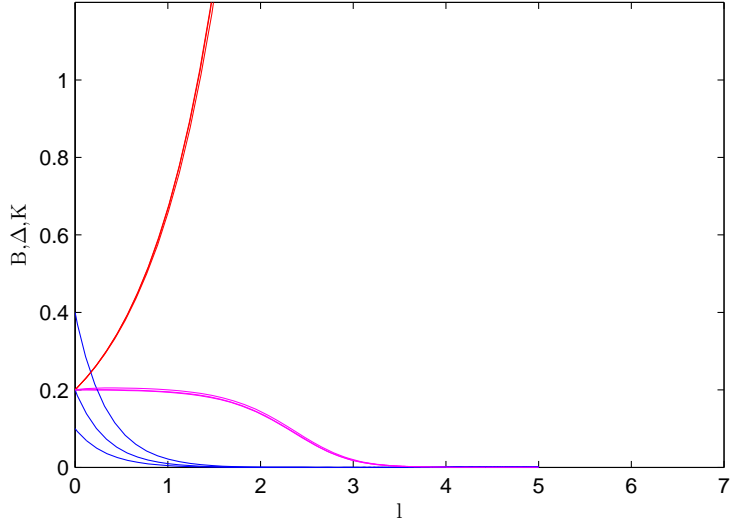}
\caption{ 
(Colour online.) This figure shows the  
variation of $\Delta, B$ and $K$ with length scale.
The curves for blue, pink and red lines are 
respectively for the $\Delta$, $K$ and $B$.  
$\Delta = 0.1, 0.2, 0.4; B=0.2, K= 0.2 $,
}
\label{Fig. 10 }
\end{figure}
\section { Berezinskii-Kosterlitz-Thouless Transition } 
Berezinskii, Kosterlitz and Thouless proposed that the disordering
is facilited by the condensation of topological defects \cite{bere,koster}.
Generally one can consider it has as a vortex. It can be
expressed as a two dimensional XY model. The orientation of
a spin is define up to an interger multiple of $ 2 \pi $.
One can consider the spin configuration in which the travel of
a closed path will see the angle rotate by $ 2 \pi n $ where $ n $ is the
topological charge.
Here we derive and discuss the appearence of Berezinskii-Kosterlitz-Thouless
(BKT) transition in our system. BKT transition is a topological quantum 
phase transition in low dimensional many body system. Here we derive the 
quantum version of it, i.e., at $ T= 0$. \\
To the best of our knowledge this is the first study to BKT transition
in interacting helical liquid to find the physics of Majorana fermion mode.\\
Before we start to discuss the appearance of BKT transition in our
system, we would like to discuss very briefly why it is necessary
to study the BKT transition. Here we study two different situations
of our model Hamiltonian. For the first case, the applied magnetic
field is absent ($B=0 $) and for the second case the proximity induced 
superconducting gap term is absent ($ \Delta =0$). 
For both of these cases only
one of the sine-Gordon coupling term is present, therefore, there is
no competition between the two mutually non local perturbation. Therefore
one can think that there is no need to study the RG to extract the
quantum phases and phase boundaries. But we still apply RG method for
the following reason. Each of these Hamiltonians consist of two part,
the first one ($H_{01}$ ) is the non-interacting where the $\phi$ and
$\theta$ fields show the quadratic fluctuations and the other part
of these Hamiltonians are the sine-Gordon coupling terms which
of either $\theta$ or $\phi $ fields. The sine-Gordon coupling term
lock the field either $\theta$ or $\phi$ in the minima of the
potential well. Therefore the system has a competition between
the quadratic part of the Hamiltonian and the sine-Gordon
coupling term and this competition will govern the low energy
physics of these Hamiltonians in different limit of the
system. 
The difference between the BKT transition and Majorana-Ising transition is
that in BKT transition one
transit from a gapped state to the gapless state (Luttinger liquid phase)
but in Majorana-Ising quantum phase transition one transit from one
gapped state to the other one. \\
For the non-interacting Helical liquid system, the system shows the
two sets of RG equations for the different limit of the
parameter space. \\
Here we derive the first set of BKT equations in absence of
applied magnetic field $(B =0)$. \\
\bea
\frac{d \Delta}{dl} ~&=&~ (2 -\frac{1}{K}) \Delta ,\non \\
\frac{dK}{dl} ~&=&~ {\Delta^2},
\label{rg2}
\eea
To reduced this equations to the standard BKT equation, we do the
following transformations.
$ \tilde{K} = 1/2K $ and  $ 1 - \tilde{K} = - y_{||} $. 
Then finally the above RG equations become, \\
\bea
\frac{d \Delta}{dl} ~&=&~ - y_{||} \Delta ,\non \\
\frac{d y_{||}}{dl} ~&=&~ - {\Delta^2},
\label{rg2}
\eea
The other set of RG equations (when $\Delta =0 $) are  
\bea
\frac{d{B}}{dl} ~&=&~ (2 - K) {B}  \non \\
\frac{dK}{dl} ~&=&~ -~ K^2  {B}^2 ~,
\label{rg2}
\eea
Here we do the following two transformations,
$ K = 2  + y_{||} $ and $ B \rightarrow B/2 $. Finally the
equation reduce the standard BKT equation. \\
\bea
\frac{d B}{dl} ~&=&~ - y_{||} B ,\non \\
\frac{d y_{||} }{dl} ~&=&~ -{B^2},
\label{rg2}
\eea
Any, BKT transition consists of three phase, the weak
coupling phase, strong coupling phase and the intermediate
coupling where the system transists from the weak coupling 
phase to strong coupling phase.
The structure of these RG equations have the general structure
for the different sets of RG equations.\\
The analysis of these RG equations consists of three different phases.\\
The first one is the weak coupling phase when ${y_{||}} > \Delta $. In this
phase there is no gapped excitation of the system, i.e, the system
is in the Luttinger liquid phase. In this phase, there is no evidence of 
Majorana fermion mode, i.e., the system is in the non-topological state.\\ 
The second one is the crossover regime, the mathematical condition for 
this crossover regime is 
$- {\Delta} < y_{||} < {\Delta} $ 
$ (- {\Delta} < (\frac{1}{2 K} -1) < {\Delta})$. 
One observe
the crossover from the weak coupling phase ($ y_{||} = |\Delta |$ )
to the strong coupling regime ($ y_{||} = - |\Delta | $). During this phase
cross over system transit from Luttinger liquid phase to the proximity
induced superconducting gap phase, i.e., $\Delta \neq 0 $. In this phase the
system has the Majorana fermion mode. For this situation, system transit
from non-topological state to the topological state.\\
The third one is the deep massive phase, the RG flows flowing off to the
strong coupling regime away from the Gaussian fixed line.\\
The structure of the second BKT equations (Eq.23 ) is the same form as that 
of the first one (Eq. 21). Therefore the analysis of these equation are the same as that
of the first one. The only difference is that here is no evidence Majorana 
fermion mode in this system. The system only shows the ferromagnetic phase.\\ 
The physics of BKT transition has studied in different condensed matter system
like in one or two dimensional superconductivity, superfludity, Josephson junction
array, melting of cryastaline thin flim, one-dimensional metal and quantum magnets.
But the explicit study of BKT transition for cavity QED lattice is absent
in the previous literature.\\
\section{Effect of Chemical Potential}
Now we consider the effect of the presence of chemical potential ($\mu $)
in the physics of interacting helical liquid. At first we consider
the absence of umklapp scattering ($ g_u$) term and the proximity
induce superconducting gap $(\Delta )$. At first we consider a
transformation, $ 2 \phi \rightarrow 2 \phi  + \delta_1 x $,
where ${\delta}_1 = - \frac{K \mu}{\pi} $. This transformation 
eliminate the term $\partial_x \phi (x) $ from the bosonized
Hamiltonian but this transformation leads to a spatially oscillating
term, i.e., $ \frac{B}{\pi} cos (\sqrt{\pi} (2 \phi + \delta_1 x ) )
= \frac{B}{\pi} cos (2 \sqrt{\pi} \phi + \delta_1 \sqrt{\pi} \phi (x))$.
In the absence of $g_u $ and $\Delta $, system shows the commensurate
to incommensuarte transition.\\
If ${\delta_1 } a >> 1$, then the term 
$ \frac{B}{\pi} cos ( \sqrt{\pi} (2 \phi + \delta_1 x ) ) $ is quickly
oscillating and average out to zero which reflect the compition 
between the $\mu $ and $B$. As a result of this compition, the RG
flow in B has to be cut-off when $ 2 \delta_1 (l) \sim 1 $. 
To the lowest order in $B, \Delta$ and $\delta_1 $, the RG flows
of $\delta_1 $ is
\beq
\frac{d \delta_1}{dl} = \delta_1 .
\eeq
When the all perturbations are relevant, they flows to the strong 
coupling phase under RG transformation. If the coupling $ B(l) $ 
reaches the strong coupling phase before $ \delta_1 (l) a $
reach to one. The phase of the system is the ferromagnetic phase. This
condition is $ 2 \delta_1 ( l^* ) a << 1 $ which translate into
the requirement $ \delta_1 (0) a <<  {B (0)}^{1/(2-K) }$.
When we consider the opposite scenario, i.e., the $B$ term 
is not able to reach the strong coupling phase but the $\Delta (l)$
term reaches the strong coupling phase and system shows the
presence of Majorana fermion mode. The presence of $\mu $ has 
no effect on it, because the $\mu $ term is related with the
$\phi$ field and the proximity induce superconductivity is
related with $\theta$ term.\\
Now we discuss the physics in presence of umklapp term, the
system posses a oscillatory term 
$\frac{g_u}{\pi} cos (4 \sqrt{\pi} \phi (x) 
+ 2 {\delta}_1 \sqrt{\pi} \phi (x))$.      
If the $g_u (l) $ term reaches the strong coupling phase earlier than
the $ B(l) $ then the system is in the ferromagnetic phase. \\
\section{ Summary and Conclusions} 
We have done the renormalization group study of interacting helical 
liquid in presence of proximity induce superconductivity and applied
magnetic field along the edge direction.  
We have found that Majorana 
fermion mode exist for this system even in the presence of repulsive
interaction and umklapp scattering. We have also studied the 
length scale dependence of $B, \Delta, K$ under the renormalization
to get the condition for the existence of Majorana fermion explicitly.
We have also studied the Berezinskii-Kosterlitz-Thouless transition.
The condition of self duality destroy in presence of umklapp scattering.\\  

%%%%%%%%%%%%%%%%%%%%%%%%%%%%%%%%%%%%%%%%%%%%%%%%%%%%%%%%%%%%%%%%%%%%
\centerline{\bf Acknowledgments}
\vskip .2 true cm
The author would like to thank the useful discussions with Prof. Dr.
A. Altland during the International School on "Topological State of Matter"
at Harish Chandra Research Institute, Allahabad. The author would also like
to thank the DST (SERB) fund. Finally, the author would like to thank, the
library of RRI for extensive support.\\

\newpage
{\bf Appendix} \\
The authors of Ref. \cite{sela} have not presented the detail derivation of the bosonized
Hamiltonian. Here we present the detail derivation of the bosonized Hamiltonian
during this derivation we follow the following references \cite{gia,senechal}.\\
In the bosonization process, one can express the fermionic field of low
dimensional quantum many body system as, \\
$ {\psi}_{R/L, \uparrow} = \frac{1}{2 \pi \alpha} {\eta_{R, \uparrow}}~
e^{i \sqrt{4 \pi} {\phi}_{R, \uparrow / \downarrow} } $.
$\eta_{L/R} $ is the Klein factor to preserve the anticommutivity of the
fermionic field which obey the Clifford algebra (here cite reference).
Here we introduce the two bosonic fields, $\theta$ and $\phi$, 
which are dual to each other. These two fields are related with the following
relations, $\phi = {\phi}_R + {\phi}_L $ and $\theta = {\theta}_R + {\theta}_L $.\\
The analytical relation between the Klein factors have mentioned in Ref. \cite{gia,senechal}. \\
%% $ \{ \eta_{L,\alpha} , \eta_{L,\beta} \}= 2 \delta_{\aplha,beta} $, 
%% $ \{ \eta_{R,\alpha} , \eta_{R,\beta} \}= 2 \delta_{\aplha,beta} $, 
%% $ \{ \eta_{R,\alpha} , \eta_{L,\beta} \}= 0 $, 

The non-interacting Hamiltonian for helical liquid is
\bea
H_0 &=& {\psi_{L,\downarrow}}^{\dagger}  (i v_F \partial_x  - \mu) {\psi_{L,\downarrow}}
+ {\psi_{R,\uparrow}}^{\dagger}  (- i v_F \partial_x  - \mu) {\psi_{R,\uparrow}} \non\\
& & = v_F [ 
{( \partial_x {\phi_{L,\downarrow}} )}^2 +
{( \partial_x {\phi_{R,\uparrow}} )}^2 ]
 - \frac{\mu}{\sqrt{\pi}} \partial_x {\phi} .  
\eea
We use the following relation during the derivation of the above Hamiltonian.
$ {\rho_{R/L , \sigma} } = \frac{1}{\sqrt{\pi}} {\partial_x } {\phi_{R/L, \sigma}} $.\\
Now we bosonize the forward scattering and umklapp scattering term, these two interaction
terms are time reversal symmetry invariant terms. \\
\bea
H_{f} & = & g_2 {\psi_{L,\downarrow}}^{\dagger} {\psi_{L,\downarrow}}
{\psi_{R,\uparrow}}^{\dagger} {\psi_{R,\uparrow}} + 
\frac{g_4}{2} [{( {\psi_{L,\downarrow}}^{\dagger} {\psi_{L,\downarrow}} )}^2 +
{( {\psi_{R,\uparrow}}^{\dagger} {\psi_{R,\uparrow}} )}^2   ] \non\\
& &
= g_2 {\rho_{L \downarrow}}{\rho_{R \uparrow}} + \frac{g_4}{2} 
[ {( \rho_{L \downarrow} )}^2 + {( \rho_{R \downarrow} )}^2 ] 
\eea  
Now we use the bosonized version of the density operator to
express the bosonized form of the above Hamiltonian. Finally the above
Hamiltonian reduced to \\
\beq
H_{f} = \frac{g_2}{\pi} {\partial_x {\phi_{L \downarrow}}} {\partial_x {\phi_{R \uparrow}}}
+ \frac{g_4}{2 \pi} [ {( \partial_x \phi_{L \downarrow} )}^2 + 
 {( \partial_x \phi_{R \uparrow} )}^2  
\eeq  
Now we consider the umklapp term, we follow the convention of Wu, Bernevig and
Zhang \cite{wu}, to write the umklapp term in a point splitted form to regularize the
theory where lattice constant use as a ultra-violet cut-off \cite{wu}. \\
\bea
H_{um} & = & - {g_u} [ {( {\psi_{L \downarrow}}^{\dagger} {\psi_{R \uparrow} } )}^2
+  {( {\psi_{R \uparrow}}^{\dagger} {\psi_{L \downarrow} } )}^2 ] \non \\
& & =  - \frac{-g_u}{ {(2 \pi)}^2 } 
[ {\eta_{L \downarrow}} {\eta_{R \uparrow}}
 {\eta_{L \downarrow}} {\eta_{R \uparrow}} 
e^{2 i \sqrt{4 \pi} ( \phi_{L \downarrow} + \phi_{R \uparrow} )} ]
 + \frac{-g_u}{ {(2 \pi)}^2 } [ {\eta_{R \uparrow}} {\eta_{L \downarrow}}
 {\eta_{R \uparrow}} {\eta_{L \downarrow}} 
e^{- 2 i \sqrt{4 \pi} ( \phi_{L \downarrow} + \phi_{R \uparrow} )} ]
\eea
Now we derive the bosonized form of the applied magnetic 
field along the edge direction.\\
\bea
H_B &=& \frac{B}{2 \pi} ( {\eta_{L, \downarrow}} {\eta_{R,\uparrow} }
e^{i \sqrt{4 \pi} ({\phi}_{L \downarrow} +{\phi}_{R \uparrow} ) + h.c)} \non\\
& & = \frac{B}{2 \pi}  {\eta_{L, \downarrow}} {\eta_{R,\uparrow} }
(e^{i \sqrt{4 \pi} \phi}  - e^{- i \sqrt{4 \pi} \phi } ) 
\eea 
\bea
H_{\Delta} & = & \frac{\Delta}{2 \pi} ( {\eta_{L, \downarrow}} {\eta_{R,\uparrow} }
e^{i \sqrt{4 \pi} (- {\phi}_{L \downarrow} +{\phi}_{R \uparrow} ) + h.c) } \non\\
& & = \frac{\Delta}{2 \pi}  {\eta_{L, \downarrow}} {\eta_{R,\uparrow} }
(e^{i \sqrt{4 \pi} \theta}  - e^{- i \sqrt{4 \pi} \theta } ) 
\eea 
Now we follow the prescription of Senechal to choose a Klein basis which
simultaneously diagonalizes all Klein factors. Here we are faceing two 
Klein factors \\
$ {\kappa}_1 =  
 {\eta_{L \downarrow}} {\eta_{R \uparrow}}
 {\eta_{L \downarrow}} {\eta_{R \uparrow}} $.
$ {\kappa}_2 =  
 {\eta_{L \downarrow}} {\eta_{R \uparrow}} $.
According to the notation of Senechal, we can write,
$ {\eta_{L \downarrow}} = 1 \otimes {\sigma}^{y}$,
$ {\eta_{R \uparrow}} = {\sigma}^{x} \otimes {\sigma}^{x}$, Therefore
one can write the $\kappa_1 $ and $\kappa_2 $ as\\
$ {\kappa_1} = \left (\begin{array}{cccc}
      -1 & 0 & 0 & 0 \\
       0 & -1 & 0 & 0 \\
       0 & 0 & -1 & 0 \\
       0 & 0 & 0 & -1
        \end{array} \right ) $ ,
$ {\kappa_2} = \left (\begin{array}{cccc}
      0 & 0 & -i & 0 \\
       0 & 0 & 0 & i \\
       -i & 0 & 0 & 0 \\
       0 & i & 0 & 0
        \end{array} \right ) .
         $
\\
The above two matrices are commute with each other, therefore one can diagonalize 
these matrices simultaneously. The eigen vectors of one matrix can be expressed
as a linear combination of eigen vectors of other matrix from which one can
construct the unitary matrix which diagonalize the both matrix $\kappa_1 $
and $\kappa_2 $ simultaneously. The diagonalize matrix, we may call it as
a $\kappa_3 $ and $\kappa_4 $.\\
$ {\kappa_3} = {U}^{-1} \kappa_1 U =  
 \left (\begin{array}{cccc}
      -1 & 0 & 0 & 0 \\
       0 & -1 & 0 & 0 \\
       0 & 0 & -1 & 0 \\
       0 & 0 & 0 & -1
        \end{array} \right ) $ ,
$ {\kappa_4} = {U}^{-1} \kappa_2 U =  
 \left (\begin{array}{cccc}
      -i & 0 & 0 & 0 \\
       0 & i & 0 & 0 \\
       0 & 0 & i & 0 \\
       0 & 0 & 0 & -i
        \end{array} \right ) $ ,

One can gauge choice by considering, $- 1$ and $ -i$ from the upper left
corner and forget the rest of the Klein space. \\
Using this convention one can write the Hamiltonian 
$H_B $ (Eq. 30) and $H_{\Delta} (Eq. 31) $ are
respectively. \\
$ H_B = \frac{B}{\pi} sin( \sqrt{4 \pi } \phi (x) ) $.\\
$ H_{\Delta} = \frac{\Delta}{\pi} sin( \sqrt{4 \pi } \theta (x) ) $.\\
$ H_{g_u} = \frac{g_u}{2 {\pi}^2} cos( \sqrt{16 \pi } \phi (x) ) $.\\

$ v = v_F + \frac{g_4}{2 \pi} $, $K = 1 - \frac{g_2}{2 \pi v_F} $. \\

Explanation of signe mismatch:\\
The author of Ref. \cite{sela} have considered 
$ \sqrt{4 \pi \phi } \rightarrow \sqrt{4 \pi \phi} + \pi/2 $,  
$ \sqrt{4 \pi \theta } \rightarrow \sqrt{4 \pi \theta} - \pi/2 $. 
Therefore the form of the full bosonized Hamiltonian is
\bea
H & = & \frac{v}{2} ( \frac{1}{K}{({\partial_x \phi })}^2 + 
K {({\partial_x  \theta })}^2  ) - \frac{\mu}{\pi} \partial_x \phi \non\\
& & + \frac{B}{\pi} cos(\sqrt{4 \pi} \phi) - \frac{\Delta}{\pi} cos(\sqrt{4 \pi \theta})
+ \frac{g_u}{2 {\pi}^2 } cos(4 \sqrt{\pi} \phi) 
\eea
\end{document}